\documentclass{aa} 
\usepackage{graphicx}
\begin{document}
\def\ffam {\hbox{$\,.\!\!^{\prime}$}}
\def\ffas {\hbox{$\,.\!\!^{\prime\prime}$}}
\def\ffM {\hbox{$\,.\!\!^{\rm M}$}}
\def\ffm {\hbox{$\,.\!\!^{\rm m}$}}
\def\ffs {\hbox{$\,.\!\!^{\rm s}$}}

\title{Discovery of water vapor megamaser emission from Mrk~1419 (NGC~2960):
       An analogue of NGC~4258?}

\author{C. Henkel\inst{1}
        \and
        J. A. Braatz\inst{2}
        \and
        L. J. Greenhill\inst{3}
        \and
        A. S. Wilson\inst{4}
        }

\offprints{C. Henkel,
\email{chenkel@mpifr-bonn.mpg.de}}

\institute{
           Max-Planck-Institut f\"ur Radioastronomie, Auf dem H\"ugel
              69, D-53121 Bonn, Germany
           \and
           National Radio Astronomy Observatory, PO Box 2, Green Bank, WV 24944, USA
           \and
           Harvard--Smithsonian Center for Astrophysics, 60 Garden St., Cambridge MA 02138, USA 
           \and 
           Department of Astronomy, Univ. of Maryland, College Park, MD 20742, USA
}

\date{Received date / Accepted date}

\abstract{Water vapor emission at 22\,GHz is reported from the nucleus of the LINER 
galaxy Mrk\,1419 (NGC~2960). Single-dish spectra of the maser source show properties 
that are similar to those seen in NGC\,4258, namely (1) a cluster of systemic 
($V$$\sim$$V_{\rm sys}$) H$_2$O features, (2) two additional H$_2$O clusters, one red- and one 
blue-shifted with respect to $V_{\rm sys}$, (3) a likely acceleration of the systemic features 
(d$V$/d$t$ = 2.8$\pm$0.5\,km\,s$^{-1}$\,yr$^{-1}$), and (4) no detectable velocity drifts 
($<$1\,km\,s$^{-1}$\,yr$^{-1}$) in the red- and blue-shifted features. Interpreting the 
data in terms of the paradigm established for NGC~4258, i.e. assuming the presence of an 
edge-on Keplerian circumnuclear annulus with the systemic emission arising from the near side
of its inner edge, the following parameters are derived: $V_{\rm rot}$ = 330--600\,km\,s$^{-1}$, 
$R$$\sim$0.13--0.43\,pc, binding mass $M$$\sim$10$^{7}$\,M$_{\odot}$, and mass density inside 
the disk $\rho$$\sim$10$^{9}$\,M$_{\odot}$\,pc$^{-3}$. With the galaxy being approximately ten 
times farther away than NGC\,4258, a comparison of linear and angular scales (the latter via 
Very Long Baseline Interferometry) may provide an accurate geometric distance to Mrk\,1419 
that could be used to calibrate the cosmic distance scale.
}

\titlerunning{The water vapor megamaser in Mrk~1419}
\authorrunning{Henkel et al.}

\maketitle


\section{Introduction}

Water vapor masers provide the only emission lines from the accretion disks of active
galactic nuclei (AGN) that can be spatially mapped. Our ability to image these objects
comes about through the very high brightness provided by the maser process and the 
fortuitous location of a line at 1.3\,cm (22\,GHz), at which wavelength we are able
to image with submilliarcsecond resolution. Very Long Baseline Interferometry (VLBI) 
images showed that, in the LINER galaxy NGC~4258, the water vapor megamaser arises in 
a thin, edge-on warped gaseous annulus between galactocentric radii of 0.16--0.28\,pc 
(Greenhill et al. \cite{greenhill95b}; Miyoshi et al. \cite{miyoshi95}; Herrnstein et al.
\cite{herrnstein99}). Maser emission is observed both near systemic velocity, arising 
from clouds at the near side of the disk, and from `satellite lines' with velocities 
$\sim$$\pm$900\,km\,s$^{-1}$ w.r.t. systemic (Nakai et al. \cite{nakai93}), arising 
from gas at the tangent points with rotational velocities directed towards and away 
from Earth. The satellite lines show an accurately Keplerian rotation curve implying a 
central mass of 3.9$\times$10$^7$\,M$_{\odot}$. The recessional velocities of the near 
systemic features are observed to be increasing at a rate of $\sim$9\,km\,s$^{-1}$\,yr$^{-1}$ 
(Haschick et al. \cite{haschick94}; Greenhill et al. \cite{greenhill95a}; Nakai et al. 
\cite{nakai95}). This increase in velocity is caused by the centripetal acceleration of 
clumps of gas in the annulus as they move across our line of sight to the central core 
(e.g. Greenhill et al. \cite{greenhill94}; Watson \& Wallin \cite{watson94}). 

In this paper we report the detection of a luminous H$_2$O megamaser in the Sa LINER galaxy
Mrk\,1419 (NGC\,2960, UGC\,05159), that appears to be quite similar to the maser in NGC\,4258.

\section{Observations}

We observed Mrk\,1419 in the $6_{16}-5_{23}$ transition of H$_2$O (rest frequency: 
22.23508~GHz) with the 100-m telescope of the MPIfR at Effelsberg\footnote{The 100-m 
telescope at Effelsberg is operated by the Max-Planck-Institut f\"ur Radioastronomie 
(MPIfR) on behalf of the Max-Planck-Gesellschaft (MPG).} on January 28 and 30, March 12,
May 7, December 15, 16, and 31, 2001, and on March 5, 2002. The beam width was 40$''$. 
The observations were made with a two channel K-band receiver in a dual beam switching 
mode with a beam throw of 2\arcmin\ and a switching frequency of $\sim$1\,Hz. System 
temperatures, including atmospheric contributions, were $\sim$100\,K (Jan, Mar, Dec 2001) 
and $\sim$150\,K (May 2001, Mar 2002) on a main beam brightness temperature scale. An
autocorrelator provided eight 40\,MHz wide bands with 512 channels each, overlapped to
produce a total bandwidth coverage of $\ga$100\,MHz with dual polarization. Flux calibration 
was obtained by measuring 3C\,286 (2.5\,Jy). Gain variations of the telescope as a 
function of elevation were taken into account and the accuracy of the calibration should 
be better than $\pm$25\%. The pointing accuracy deduced from measurements of 0851+202, 
0906+01, 3C\,218 and 1127--14 was better than 10$''$.

\section{Results}

\begin{figure}[t]
\centering
\resizebox{12.0cm}{!}{\rotatebox[origin=br]{-90}{\includegraphics{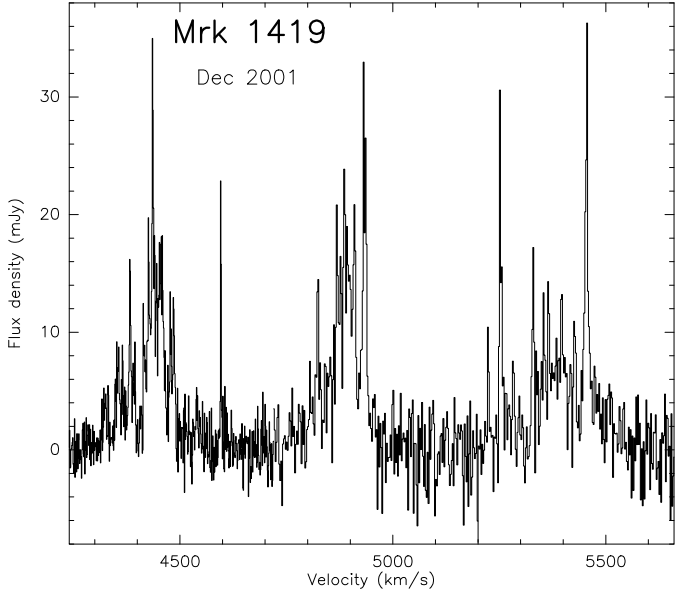}}}
\caption{Averaged 22\,GHz H$_2$O spectrum of Mrk\,1419 ($\alpha_{2000}$ = 9$^{\rm h}$40$^{\rm m}$36\ffs5, 
$\delta_{2000}$ = 3$^{\circ}$34$'$38$''$) including data taken on Dec. 15, 16, and 31. 
The velocity scale is with respect to the Local Standard of Rest (LSR) and uses the optical 
convention that is equivalent to c$z$ ($V_{\rm sys}$ = 4932\,km\,s$^{-1}$; Braatz et al. 
\cite{braatz97}). Channel spacings are 2.18\,km\,s$^{-1}$ for the systemic 
and red-shifted and 1.09\,km\,s$^{-1}$ for the blue-shifted features.
\label{fig1}}
\end{figure}

\begin{figure}[t]
\centering
\resizebox{12.0cm}{!}{\rotatebox[origin=br]{-90}{\includegraphics{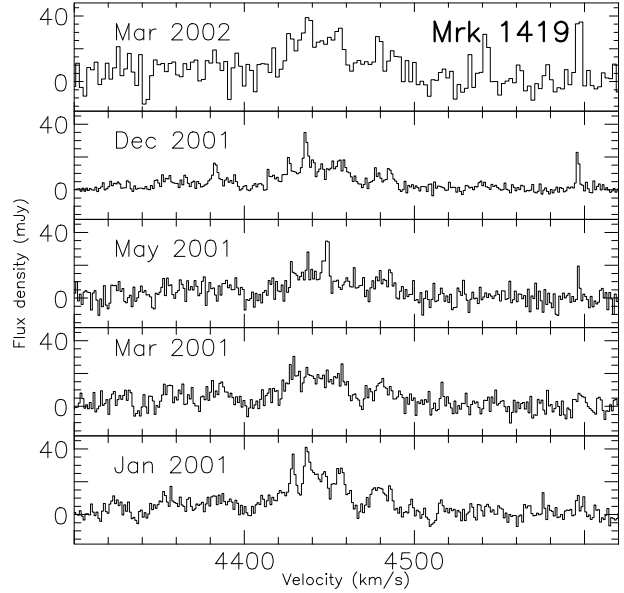}}}
\caption{Blue-shifted H$_2$O features of Mrk\,1419. Channel spacings are 2.18\,km\,s$^{-1}$ 
for the upper panel and 1.09\,km\,s$^{-1}$ for the lower panels.
\label{fig2}}
\end{figure}

\begin{figure}[t]
\centering
\resizebox{12.0cm}{!}{\rotatebox[origin=br]{-90}{\includegraphics{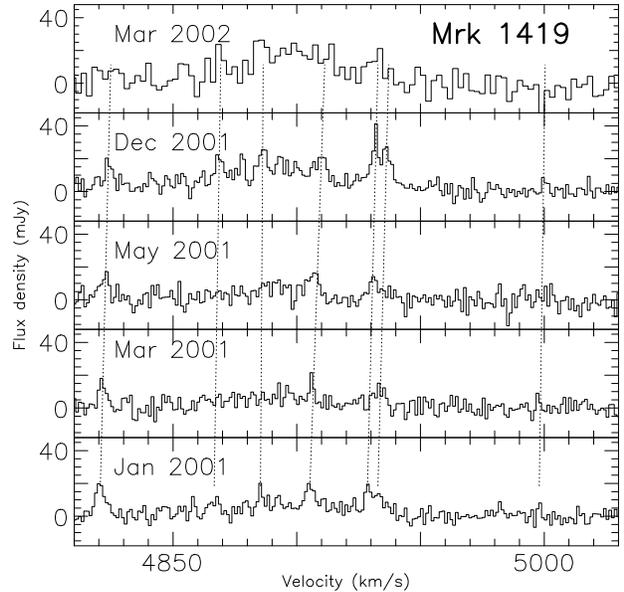}}}
\caption{Near systemic H$_2$O features of Mrk\,1419. Dotted lines connect features 
detected both in Jan. and Dec. 2001 (for Gaussian fits, see Table 1). Channel spacings 
are 2.18 for the upper panel and 1.09\,km\,s$^{-1}$ for the lower panels.
\label{fig3}}
\end{figure}

\begin{figure}
\centering
\resizebox{12.0cm}{!}{\rotatebox[origin=br]{-90}{\includegraphics{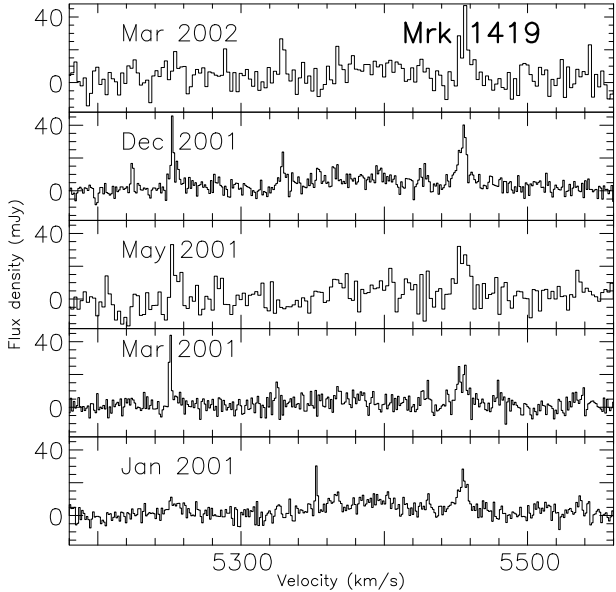}}}
\caption{Red-shifted H$_2$O features of Mrk\,1419. Channel spacings are 2.18 for the 
May 2001 and March 2002 spectra and 1.09\,km\,s$^{-1}$ for the other profiles.
\label{fig4}}
\end{figure} 

Fig.\,\ref{fig1} shows a characteristic 22\,GHz H$_2$O spectrum of Mrk1419, displaying 
the velocity interval from 4240 to 5660\,km\,s$^{-1}$. There are three groups of maser 
features: The blue-shifted components are found between 4320 and 4600\,km\,s$^{-1}$, 
the red-shifted ones between 5200 and 5550\,km\,s$^{-1}$, and the systemic ones between 
4820 and 5000\,km\,s$^{-1}$. Systemic line intensities are consistent with the upper limit 
provided by Braatz et al. (\cite{braatz96}). The two gaps devoid of line emission 
have widths of 200--220\,km\,s$^{-1}$. Figs.\,\ref{fig2}--\ref{fig4} show 22\,GHz H$_2$O 
spectra from all five epochs with an enlarged velocity scale. Data obtained within 
a month do not show significant differences and were averaged. Integrated fluxes of our most 
sensitive spectra, those taken during January and December 2001, are about 1.8 and 1.25 
(blue-shifted), 0.75 and 1.55 (systemic) and 1.2 and 1.6\,Jy\,km\,s$^{-1}$ (red-shifted), 
respectively, implying total (isotropic) 22\,GHz H$_2$O luminosities of $\sim$350 and 
$\sim$425\,L$_{\odot}$ at these times (assumed distance: $D$ = 65\,Mpc). Not accounting 
for (possibly small) beaming angles, Mrk\,1419 is thus one of the more luminous known 
H$_2$O megamasers. Some individual components show strong variability. The feature at 
$\sim$5250\,km\,s$^{-1}$ varies in intensity by a factor $\ga$5; the $\sim$5352.5\,km\,s$^{-1}$ 
component (Fig.\,\ref{fig4}) was only detected in January 2001.

On January 28, 2001, we searched for maser emission in a wide velocity range, from
3430 to 6500\,km\,s$^{-1}$. No significant very blue or red feature was detected at 
noise levels of 7--10\,mJy and channel spacings of 1.09\,km\,s$^{-1}$. 

\begin{table}
\caption{H$_2$O in Mrk\,1419: Radial velocities of systemic (left) and non-systemic 
(right) components detected both in January and December 2001. \label{fittare}}
\begin{center}
\begin{tabular}{cc|cc}
\\
\multicolumn{4}{c}{Epoch} \\
2001.08         &   2001.97       & 2001.08        &   2001.97        \\
\hline
4820.6$\pm$0.3  &  4823.9$\pm$0.3 &  4437.0$\pm$0.2  & 4436.2$\pm$0.2 \\
4866.6$\pm$0.8  &  4868.6$\pm$0.5 &  4456.6$\pm$0.5  & 4456.6$\pm$0.7 \\
4885.3$\pm$0.3  &  4886.2$\pm$0.4 &  4597.4$\pm$0.6  & 4596.1$\pm$0.1 \\
4905.3$\pm$0.3  &  4910.0$\pm$0.5 &        --        &       --       \\
4928.6$\pm$0.3  &  4931.9$\pm$0.1 &        --        &       --       \\
4932.8$\pm$0.9  &  4936.3$\pm$0.3 &  5253.1$\pm$1.1  & 5252.1$\pm$0.1 \\
4998.0$\pm$0.2  &  5001.1$\pm$0.4 &  5455.0$\pm$0.4  & 5454.9$\pm$0.3 \\
\hline
\end{tabular}
\end{center}
\end{table}

\section{Discussion}

A comparison of Mrk\,1419 with the best studied H$_2$O megamaser source, NGC\,4258, shows 
a striking similarity: in both galaxies, there are systemic features as well as red-, and 
blue-shifted groups of H$_2$O components that symmetrically bracket $V_{\rm sys}$. 
Furthermore, accounting for a ratio of $\sim$10 in recessional velocity and distance, 
the systemic features (a few Jy towards NGC\,4258, a few 10\,mJy towards Mrk\,1419) show 
similar luminosities. An interpretation in terms of the NGC\,4258-paradigm (H$_2$O
emission from a Keplerian circumnuclear disk; see Sect.\,1) is therefore suggestive. 
Detection of acceleration of the systemic features, caused by centripetal acceleration 
at the front or back side of the putative edge-on disk, and the absence of such a drift 
in the non-systemic features from the tangentially seen parts of the disk would provide 
independent evidence and would go a long way towards confirming an analogous model to 
that already established for NGC\,4258.
 
\begin{figure*}
\centering
\vspace{-1.5cm}
\resizebox{16.0cm}{!}{\rotatebox[origin=br]{-90}{\includegraphics{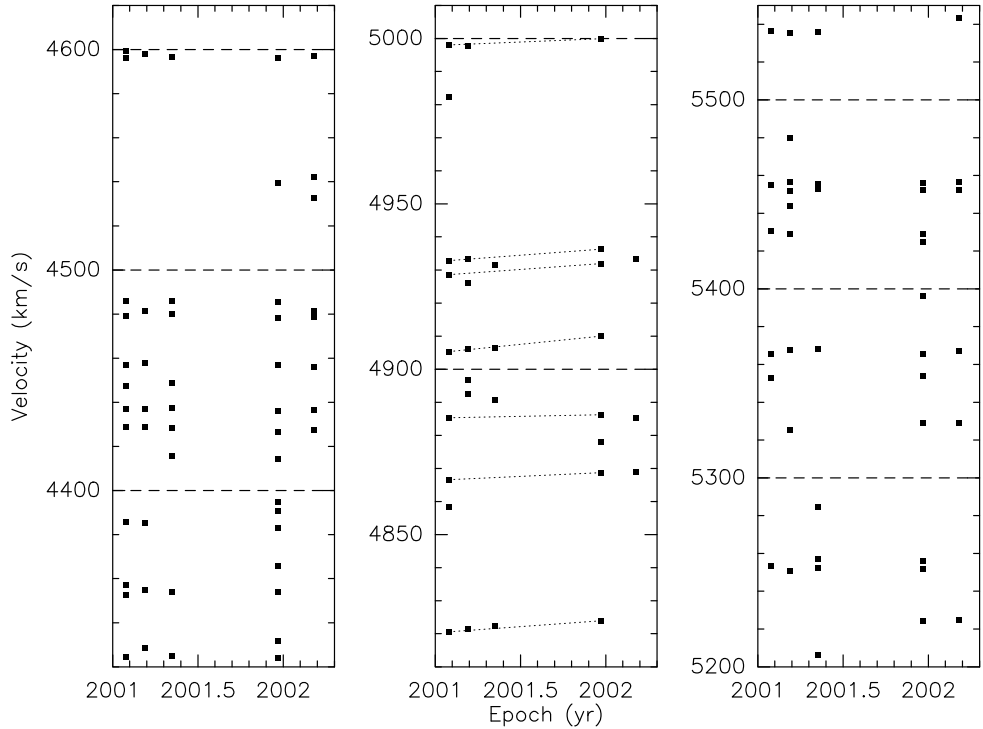}}}
\caption{Velocities of individual H$_2$O features obtained from Gaussian fits with standard 
deviations of 0.1--1.8\,km\,s$^{-1}$. From left to right the blue-shifted, systemic, and 
red-shifted components of Mrk\,1419 are shown as a function of time. Dotted lines (central 
panel) connect systemic features detected in both Jan. and Dec. 2001 (see Sect.\,4). 
\label{fig5}}
\end{figure*} 

Fig.\,\ref{fig5} shows radial velocities for blue, systemic, and red features (as shown 
in Figs.\,\ref{fig2}--\ref{fig4}, respectively) as a function of time. Our most sensitive 
spectra are from Jan. and Dec. 2001. Connecting systemic features observed during both 
epochs (dotted lines in Fig.\,\ref{fig5}, central panel) a positive drift is detected 
in each of a total of seven components. This is also shown in Fig.\,\ref{fig3}, where 
dotted lines connect the line peaks obtained from Gaussian fits. Velocities and standard 
deviations of these components are given in Table 1. Fits to the data lead to an 
acceleration of 2.8$\pm$0.5\,km\,s$^{-1}$\,yr$^{-1}$, the error being the standard 
deviation of the mean. Assuming a normal error distribution, the likelihood of an acceleration
$<$1.0\,km\,s$^{-1}$\,yr$^{-1}$ is $\sim$10$^{-4}$. For all seven components the measured 
acceleration is larger than the sum of the related standard deviations (the scatter can be 
explained by groups of individual maser features with varying intensities (e.g. Genzel \& Downes 
\cite{genzel77})). The corresponding drift for the red- and blue-shifted components 
(--0.6$\pm$0.3\,km\,s$^{-1}$) is not significant and the likelihood for an overall acceleration
within $|\dot{V}_{\rm drift}|$$<$1\,km\,s$^{-1}$\,yr$^{-1}$ is $\sim$90\%.

Having demonstrated that the systemic features likely show a secular acceleration, while the 
red- and blue-shifted components lack a similar drift, it makes sense to assume that the 
maser emission in Mrk\,1419 arises from an almost edge-on circumnuclear annulus like that 
in NGC\,4258. Assuming similar morphologies for NGC\,4258 and Mrk\,1419, no interferometric 
data are needed to estimate the size of the disk and the mass enclosed (for equations, 
see e.g. Ishihara et al. \cite{ishihara01}). Assuming that the systemic emission arises (as in 
NGC\,4258) from the near side of the inner edge of the annulus, where Keplerian velocities 
are highest, the difference in velocity between systemic and red- and blue-shifted features 
(see Figs.\,\ref{fig2}--\ref{fig4}) gives $V_{\rm rot,in}$ = 600$\pm$20\,km\,s$^{-1}$. 
$V_{\rm rot,in}$, the velocity drift mentioned above, and an assumed inclination of 
90$^{\circ}$ then yield a radius $R_{\rm in}$ = 0.13$\pm$0.03\,pc for the inner edge of 
the disk ($R$$\propto$$V_{\rm rot}^2$, so the disk would be smaller if the systemic features 
do not arise from the inner edge but farther out). For the outer edge with $V_{\rm rot,out}$ 
$\sim$ 330$\pm$20\,km\,s$^{-1}$ (Figs.\,\ref{fig2}--\ref{fig4}), we then find $R_{\rm out}$ 
= ($V_{\rm rot,in}/V_{\rm rot,out}$)$^{2}$ $\times$ $R_{\rm in}$ $\sim$ 0.43$\pm$0.08\,pc. 
The full angular extent of the disk should thus not greatly exceed 2\,mas ($D$=65\,Mpc). 
Still assuming that the systemic features arise from the inner edge, the binding mass and 
the mass density encircled by the disk become 1.1$\pm$0.3$\times$10$^{7}$\,M$_{\odot}$ and 
1.2$\pm$0.4\,$\times$10$^{9}$\,M$_{\odot}$\,pc$^{-3}$, respectively.

\section{Concluding remarks}

All four galaxies with (putative) circumnuclear H$_2$O disks and reported velocity drift
of the systemic emission (NGC\,4258, Mrk\,1419, NGC2639 (Wilson et al. \cite{wilson95}), 
and IC\,2560 (Ishihara et al. \cite{ishihara01})) show galactic stellar disks seen at 
intermediate inclination. The (presumably) edge-on nuclear disks are therefore 
substantially inclined relative to the plane of the parent galaxies as has also been
inferred for Seyfert galaxies from the orientation of their nuclear radio continuum ejecta
(e.g. Ulvestad \& Wilson \cite{ulvestad84}). None of the sources shows negative acceleration, 
a characteristic of emission from the far side of the disk. Although there may be no masers
there, symmetry arguments suggest that instead, absorption in the vicinity of the central engine
blocks our view toward the back side of the circumnuclear disk. That red, blue, and systemic 
features have similar intensities is unique to Mrk\,1419, but it is not yet possible to judge 
why the source is different. Mrk\,1419 should become an important target for a direct determination 
of the geometric distance of a galaxy approximately ten times as distant as NGC\,4258 (for 
NGC\,4258, see Herrnstein et al. \cite{herrnstein99}, Maoz et al. \cite{maoz99}). A measurement 
of the distance requires detailed mapping of the structure of the disk via VLBI. {\it If} a 
simple disk model can be established, the resulting distance could then be used to check 
calibration of more common optical or near infrared distance indicators, thus helping to 
establish the cosmic distance scale reliant at least in part on geometric arguments.

\begin{acknowledgements}
This research was supported by the National Science Foundation through grant AST\,9527289 to the 
University of Maryland.
\end{acknowledgements}

\end{document}